\documentclass[10pt]{extarticle}

\usepackage[utf8]{inputenc}
\usepackage[T1]{fontenc}
\usepackage{lmodern}
\usepackage[a4paper,margin=0.82in]{geometry}
\usepackage{setspace}
\usepackage{microtype}
\usepackage{hyperref}
\usepackage{graphicx}
\usepackage{booktabs}
\usepackage{array}
\usepackage{enumitem}
\usepackage{amsmath}
\usepackage{url}
\usepackage{float}
\usepackage{xcolor}
\usepackage{caption}
\usepackage{tabularx}
\usepackage{abstract}
\usepackage{longtable}

\newcolumntype{Y}{>{\raggedright\arraybackslash}X}
\newcolumntype{L}[1]{>{\raggedright\arraybackslash}p{#1}}

\setlist[itemize]{leftmargin=1.4em, topsep=2pt, itemsep=2pt, parsep=0pt, partopsep=0pt}
\setlist[enumerate]{leftmargin=1.5em, topsep=2pt, itemsep=2pt, parsep=0pt, partopsep=0pt}

\hypersetup{
colorlinks=true,
linkcolor=blue,
citecolor=blue,
urlcolor=blue,
pdftitle={Balanced Workforce: Governance-by-Design for Privacy-Preserving Inter-Firm Workforce Leasing},
pdfauthor={Melody Amaizu, Martin Stojkovski, Ariton Verush},
pdfkeywords={socio-technical systems, governance by design, workforce leasing, digital labor platforms, enterprise architecture, business model canvas, e3value, privacy by design, ERP integration}
}

\begin{document}

\begin{center}
\vspace{1.85cm}

{\fontsize{20}{24}\selectfont\bfseries
Balanced Workforce: Governance-by-Design for\\
Privacy-Preserving Inter-Firm Workforce Leasing
\par}

\vspace{1.55cm}

{\large\itshape
A Socio-Technical Architecture, Business Model, and Risk Framework\\
for a Talent-Sharing Platform
\par}

\vspace{1.55cm}

{\normalsize
\textbf{Melody Amaizu \quad Martin Stojkovski \quad Ariton Verush}\\[0.35em]
MSc in Computer Science\\
University of Bern\\
Bern, Switzerland\\[0.45em]

\vspace{1.25cm}

\begin{tabular}{c}
\texttt{melody.amaizu@students.unibe.ch}\\
\texttt{stojkovski.martin1@gmail.com}\\
\texttt{ariton.verush@gmail.com}
\end{tabular}
\par}

\vspace{1.25cm}

{\normalsize
Design of Governance in Socio-Technical Information Systems seminar report: June 2024\\
Revised and expanded as a governance framework paper: June 2026
\par}

\end{center}

\vspace{0.6cm}

\begin{abstract}
Workforce demand is uneven across organizations. Project-based companies may simultaneously face skill shortages in one unit while other firms hold underutilized employees with relevant expertise. Conventional hiring, contracting, and temporary agency models address parts of this problem, but they also create legal, ethical, organizational, and data-governance risks. This paper reframes a seminar project called \emph{Balanced Workforce} into a governance-by-design framework for privacy-preserving inter-firm workforce leasing. The proposed Balanced Workforce Leasing Service (BWLS) enables companies to list temporary talent availability, discover anonymized skill profiles, negotiate assignments, and document agreements through locally deployed connectors and a minimal central coordination layer. The framework combines socio-technical governance, enterprise architecture, business model design, e3value-based value exchange modeling, and privacy-by-design principles. The paper presents the system concept, stakeholder model, process phases, architecture, business model, value network, and legal/ethical/operational risk analysis. It argues that workforce leasing platforms should not be designed only as marketplaces. They require consent mechanisms, traceability, role-based access control, data minimization, contractual safeguards, dispute handling, and institutional accountability. The contribution is a structured framework and design artifact for future research on governed digital labor infrastructures. The paper does not claim deployment results or empirical validation; instead, it provides a publishable design framework that can be evaluated through expert review, stakeholder workshops, prototype testing, and regulatory analysis.
\end{abstract}

\textbf{Keywords:} socio-technical systems; governance by design; workforce leasing; digital labor platforms; enterprise architecture; business model canvas; e3value; privacy by design; ERP integration; auditability.

\newpage

\section{Introduction}

Organizations increasingly operate under fluctuating demand, project-based workloads, platform-mediated coordination, and rapidly changing skill requirements. A company may urgently need specialized employees for a short-term project, while another company may temporarily have qualified employees with no active assignment. Conventional hiring, outsourcing, contracting, and temporary agency arrangements address parts of this problem, but they can be slow, costly, legally complex, or insufficiently adapted to peer-to-peer inter-firm talent sharing \cite{EU2008TemporaryAgencyWork,DeStefano2016JustInTime,Kellogg2020AlgorithmsAtWork}.

The \emph{Balanced Workforce} concept addresses this problem by exploring a governed platform for temporary inter-firm employee leasing. The central idea is simple: if Company 1 has an employee who is temporarily underutilized and voluntarily available, and Company 2 has a temporary shortage of matching skills, a governed digital infrastructure can help the companies discover the match, negotiate terms, record agreements, and preserve traceability. However, this apparently simple mechanism is socio-technically complex. It involves employers, employees, enterprise systems, platform operators, legal agreements, privacy constraints, intellectual property, cross-company trust, and worker autonomy. Socio-technical systems research emphasizes that such systems cannot be understood as purely technical artifacts because organizational outcomes emerge from interactions among people, technologies, institutions, work practices, and governance arrangements \cite{Baxter2011SocioTechnical,Mumford2006Story}.

This paper presents the Balanced Workforce Leasing Service (BWLS) as a governance-by-design framework. Rather than proposing an unregulated labor marketplace, the framework embeds governance mechanisms into the platform architecture itself. The design uses locally deployed company connectors, anonymized skill listings, minimal central coordination, secure communication, contract management, audit logging, and role-based access. The aim is to make temporary workforce sharing technically feasible while preserving legal compliance, employee consent, and organizational accountability. This framing follows platform-governance, enterprise-architecture, and privacy-engineering perspectives, where access control, traceability, data minimization, and accountability are treated as first-order design decisions rather than after-the-fact policy additions \cite{Tiwana2010PlatformEvolution,deReuver2018DigitalPlatforms,OpenGroup2022ArchiMate,GDPR2016,EDPB2020DPbDD,Spiekermann2009EngineeringPrivacy}.

The work builds on a seminar project in \emph{Design of Governance in Socio-Technical Information Systems}. The original project produced a final report, design diagrams, process models, business model canvas elements, e3value value-exchange analysis, architecture sketches, and a risk assessment. This paper reframes that material as a structured research-style framework suitable for public preprint release. It combines design-science framing, socio-technical analysis, enterprise architecture, business model analysis, value-exchange modeling, and risk analysis \cite{Hevner2004DesignScience,Gregor2006DesignScience,Osterwalder2010BusinessModel,Gordijn2001E3Value}. It is intentionally positioned as a design and framework paper, not as a deployed empirical evaluation.

\subsection{Research Questions}

The paper is guided by four research questions:

\begin{itemize}
\item \textbf{RQ1:} How can an inter-firm workforce leasing platform be designed as a socio-technical system rather than only as a digital marketplace?
\item \textbf{RQ2:} Which governance mechanisms are required to protect employee consent, privacy, traceability, legal compliance, and organizational accountability?
\item \textbf{RQ3:} How can enterprise architecture, connector-based deployment, and business model design support a scalable workforce leasing service for ERP and non-ERP companies?
\item \textbf{RQ4:} What legal, ethical, and operational risks must be addressed before such a platform can be responsibly evaluated or deployed?
\end{itemize}

\subsection{Contributions}

This paper makes five contributions:

\begin{itemize}
\item it reframes a seminar-based Balanced Workforce project into a governance-by-design framework for inter-firm workforce leasing;
\item it defines a socio-technical system architecture based on local company connectors, minimal central coordination, anonymized listings, secure negotiation, and audit logging;
\item it connects the architecture to business modeling through a business model canvas and e3value-inspired value exchange analysis;
\item it provides a structured legal, ethical, and operational risk matrix with corresponding governance controls;
\item it proposes an evaluation agenda for future expert review, stakeholder validation, prototype testing, and regulatory assessment.
\end{itemize}

The broader argument is that digital platforms for workforce leasing must be governed at the level of architecture, process, and institutional rules. A platform that only optimizes matching efficiency may create unacceptable risks. A platform that embeds governance from the start can become a more responsible basis for experimentation.

\section{Background and Related Work}

\subsection{Socio-Technical Systems and Governance-by-Design}

Socio-technical systems research emphasizes that organizational outcomes emerge from the interaction between people, technologies, institutions, practices, and environments \cite{Baxter2011SocioTechnical,Mumford2006Story}. Information systems are therefore not neutral technical artifacts; they configure work, roles, responsibilities, and power relations. For workforce leasing, this means that architecture choices such as where profiles are stored, how consent is recorded, and who can inspect transaction logs directly affect trust and accountability.

Governance-by-design extends this idea by treating governance mechanisms as part of system design rather than as external documentation added after deployment \cite{Tiwana2010PlatformEvolution,deReuver2018DigitalPlatforms}. In BWLS, governance is implemented through consent checkpoints, role-based access, local data storage, minimal central data, audit trails, contract workflows, dispute records, and policy-aware integration with enterprise systems. These mechanisms translate broad principles such as transparency and accountability into concrete system functions.

\subsection{Temporary Agency Work and Digital Labor Platforms}

Temporary agency work is usually understood as a triangular employment relationship involving a worker, an agency or employer, and a client organization \cite{EU2008TemporaryAgencyWork,DeStefano2016JustInTime}. European regulation emphasizes worker protection and equal treatment for temporary agency workers, while also recognizing that temporary work can provide flexibility for companies. BWLS differs from standard temporary agency work because it focuses on inter-firm leasing of employees who remain attached to their original employer. However, it inherits similar legal and ethical concerns: working conditions, consent, supervision, payment responsibility, liability, and dispute resolution.

Digital labor platforms add another layer of complexity \cite{Wood2019GoodGigBadGig,Kellogg2020AlgorithmsAtWork,deReuver2018DigitalPlatforms}. Platforms can reduce search costs and improve matching, but they may also increase surveillance, asymmetry, algorithmic management, and opacity. A workforce leasing platform must therefore avoid treating employees as anonymous commodities. Employee autonomy, explanation, appeal, and control over shared information are central governance requirements \cite{Nissenbaum2004Privacy,Spiekermann2009EngineeringPrivacy}.

\subsection{Business Model and Value Exchange Modeling}

The Balanced Workforce concept is not only a technical artifact. It also requires a sustainable business model. The Business Model Canvas is useful for mapping customer segments, value propositions, channels, key resources, partnerships, costs, and revenue streams \cite{Osterwalder2010BusinessModel}. e3value modeling complements this by emphasizing economic value exchanges among actors in a network, such as companies with surplus talent, companies needing talent, employees, ERP providers, and the platform operator \cite{Gordijn2001E3Value}.

In this paper, these models are used not as final financial validation, but as analytical tools. They clarify who receives value, who pays, who carries risk, and which relationships require governance. This is especially important because a workforce leasing service may generate revenue from licenses, subscriptions, transactions, integration services, and premium support while also imposing compliance costs and ethical responsibilities.

\subsection{Enterprise Architecture, Privacy, and Data Protection}

Enterprise architecture methods such as ArchiMate support the description of relationships among business processes, application services, data flows, and technology layers \cite{OpenGroup2022ArchiMate,ISO42010}. This is useful for BWLS because the platform spans organizations and must integrate with HR systems, ERP systems, authentication infrastructure, and communication channels.

Privacy and data protection are central to the design. Under data protection principles such as lawfulness, fairness, transparency, purpose limitation, data minimization, storage limitation, integrity, confidentiality, and accountability, a workforce leasing platform should avoid unnecessary personal-data centralization \cite{GDPR2016,EDPB2020DPbDD}. Article 25 of the GDPR also motivates data protection by design and by default \cite{GDPR2016,EDPB2020DPbDD,Spiekermann2009EngineeringPrivacy}. BWLS therefore uses anonymized listings, local profile control, access logging, and role-based permissions as first-order architectural commitments rather than optional features.


\section{Method and Design Context}

This paper is based on a design-oriented reinterpretation of a seminar project. The original material included a final report, working documents, blackboard sketches, a system architecture diagram, business model notes, an e3value model, and an analysis of legal, ethical, and operational issues. The present paper reorganizes that material into a publishable framework contribution.

The method follows a design-science logic: first, a practical problem is identified; second, an artifact is proposed; third, the artifact is analyzed through business, architecture, and governance lenses; and fourth, its limitations and evaluation needs are made explicit \cite{Hevner2004DesignScience,Gregor2006DesignScience}. In this paper, the artifact is a governed workforce leasing platform with connectors, local deployment, anonymized listings, secure negotiation, and auditability.

The paper does not report a deployed system, a user study, or measured business performance. The evidence is conceptual and design-based. This positioning is important for responsible publication: the paper contributes a framework and design agenda, not proof that the platform improves employment outcomes or complies automatically with every jurisdiction.

\begin{figure}[H]
\centering
\includegraphics[width=0.9\textwidth]{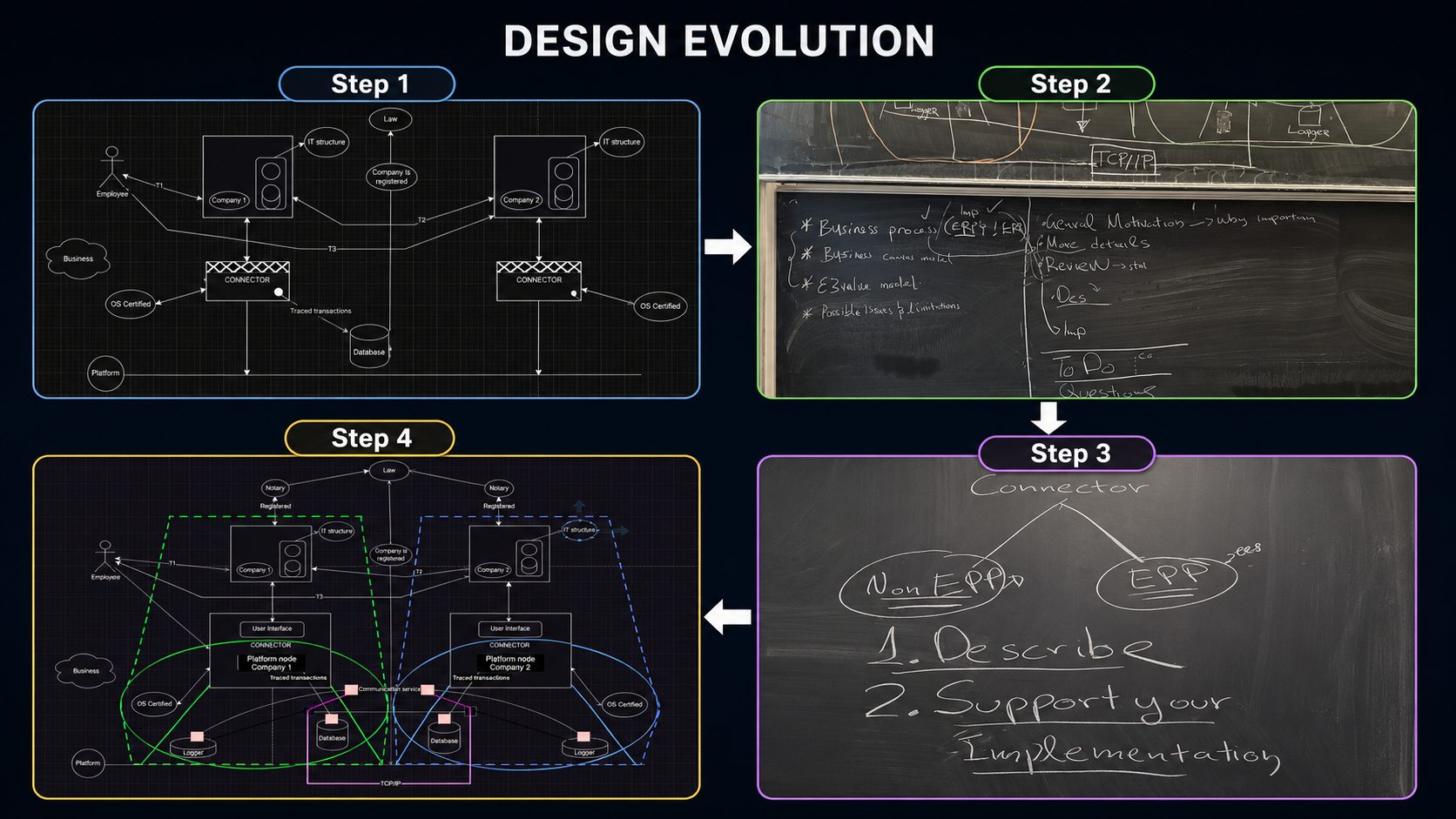}
\caption{Seminar design artifact showing the early conceptual split between ERP and non-ERP organizations and the role of a connector layer. The image is included as design-process evidence rather than as a final architecture.}
\label{fig:design_process}
\end{figure}


\section{System Concept: Balanced Workforce Leasing Service}

The BWLS system concept combines ideas from temporary workforce coordination, platform governance, and enterprise architecture. Unlike ordinary labor-matching platforms, the system must represent companies, employees, enterprise systems, legal actors, and platform services as interacting socio-technical components \cite{EU2008TemporaryAgencyWork,DeStefano2016JustInTime,Tiwana2010PlatformEvolution,OpenGroup2022ArchiMate}.

\subsection{Stakeholders}

BWLS connects several stakeholder groups. Companies with surplus talent act as lessors. Companies with temporary demand act as lessees. Employees are the human resources whose skills and time may be leased, but only with explicit participation and informed consent. IT departments and technology providers maintain the connectors, integrations, security controls, and deployment environment. Legal and compliance actors define contract templates, labor-law constraints, data-processing rules, and dispute procedures. ERP providers may participate as integration partners.

\begin{table}[H]
\centering
\caption{Core stakeholders in the Balanced Workforce Leasing Service}
\label{tab:stakeholders}
\small
\begin{tabularx}{\textwidth}{L{3.4cm}Y}
\toprule
\textbf{Stakeholder} & \textbf{Role and interest} \\
\midrule
Company 1 / lessor & Lists temporarily available employees, reduces idle workforce costs, preserves employment continuity, and maintains oversight of employee participation. \\
\addlinespace
Company 2 / lessee & Searches for short-term skills, negotiates temporary assignments, and avoids long-term hiring when demand is temporary. \\
\addlinespace
Employee & Controls participation, shared profile information, assignment acceptance, feedback, history, and personal terms. \\
\addlinespace
BWLS platform operator & Provides connectors, coordination services, licensing, support, audit infrastructure, and governance workflows. \\
\addlinespace
ERP and HR system providers & Integrate BWLS connectors into existing enterprise systems and may receive value from additional functionality. \\
\addlinespace
Legal/compliance actors & Ensure that contracts, labor conditions, data processing, liability, and dispute handling are valid in the relevant jurisdiction. \\
\bottomrule
\end{tabularx}
\end{table}

\subsection{Process Phases}

The platform is organized around four process phases. The labels T0--T3 are used to make the process auditable and easy to model, following the broader design goal of translating governance requirements into explicit system states and traceable workflow events \cite{Tiwana2010PlatformEvolution,Gregor2006DesignScience}.

\begin{itemize}
\item \textbf{T0 - Integration and activation:} a company installs a connector or standalone application, configures local access, and activates its license or subscription.
\item \textbf{T1 - Talent listing:} employees express interest, HR verifies eligibility, and approved non-sensitive skill profiles are listed with employee consent.
\item \textbf{T2 - Talent matching and negotiation:} a lessee company searches anonymized availability records, expresses interest, and begins traceable negotiation with the lessor.
\item \textbf{T3 - Final agreement and deployment:} personal terms, contractual details, signatures, assignment responsibilities, and documentation are finalized and logged.
\end{itemize}

The T2 phase is especially governance-sensitive because it connects companies and may trigger information exchange. At this point, the platform must preserve anonymity until appropriate, record requests, enforce access rights, and ensure that communication channels remain auditable.

\subsection{ERP and Non-ERP Access Models}

A central architectural decision in BWLS is that participation should not depend on a single type of enterprise infrastructure. The original design therefore distinguishes between two access pathways: an ERP-integrated model for organizations with mature enterprise or HR systems, and a lightweight non-ERP model for smaller organizations or companies with limited digital infrastructure. This distinction is important because workforce leasing is not only a technical matching problem. It is also an adoption problem. A platform restricted to large companies with advanced ERP environments would exclude many small and medium-sized enterprises, even though such organizations may frequently experience workload fluctuations, project-specific skill gaps, or temporary underutilization of employees.

In the ERP-integrated model, BWLS is accessed through a connector installed into an existing enterprise system such as SAP, Oracle, Microsoft Dynamics, Salesforce, or a comparable HR or resource-planning platform. The connector acts as a controlled boundary between the internal company environment and the broader BWLS coordination layer. Employee profile data, availability, approval status, and leasing requests can be mapped from internal HR structures to the platform's standardized representation. This model has the advantage of fitting into existing organizational workflows: HR departments can approve listings inside familiar systems, identity management can rely on enterprise authentication mechanisms, and transaction logs can be aligned with internal compliance procedures. However, this model also creates stronger engineering requirements around API security, data mapping, version compatibility, permission management, and integration testing.

The non-ERP model serves organizations that do not operate a sophisticated ERP or HR system. In this pathway, companies may use a standalone web application, a simplified desktop or web portal, or a spreadsheet-compatible plugin. The goal is to preserve the core BWLS functions---profile creation, skill listing, availability declaration, approval, search, request handling, negotiation support, and final agreement documentation---without requiring expensive enterprise infrastructure. This model lowers the adoption barrier but shifts responsibility toward usability, guided setup, safe default configurations, clear documentation, and platform-side enforcement of traceability.

From a software-engineering perspective, this dual-access strategy increases architectural complexity but improves inclusiveness and scalability. It allows BWLS to support heterogeneous organizations while maintaining a common transaction model. The platform can standardize core events such as listing creation, employer approval, talent search, request submission, negotiation start, contract upload, agreement confirmation, deployment, and post-assignment feedback. Whether these events originate from an ERP connector or a lightweight portal, they can be represented as traceable transactions in the platform's audit layer.

\begin{table}[H]
\centering
\caption{ERP and non-ERP access models}
\label{tab:access_models}
\small
\begin{tabularx}{\textwidth}{L{3.0cm}Y Y}
\toprule
\textbf{Dimension} & \textbf{ERP-integrated model} & \textbf{Non-ERP model} \\
\midrule
Target organizations & Medium and large companies with mature ERP, HR, or resource-planning systems & Small and medium-sized organizations using lightweight digital tools \\
\addlinespace
Access mechanism & Certified connector integrated into SAP, Oracle, Microsoft Dynamics, Salesforce, or comparable systems & Standalone web application, simplified portal, desktop tool, or spreadsheet-compatible plugin \\
\addlinespace
Primary goal & Embed BWLS into existing enterprise workflows and compliance procedures & Lower the adoption barrier while preserving governance and traceability \\
\addlinespace
Authentication and authorization & Enterprise identity management, single sign-on, LDAP, and role-based permissions & Platform-managed accounts, secure login, simplified permissions, and safe defaults \\
\addlinespace
Traceability & Requests, approvals, negotiations, and agreements recorded through enterprise-integrated logs and BWLS audit events & Actions recorded inside the portal, with mandatory agreement upload or confirmation \\
\addlinespace
Main challenge & API security, connector certification, data mapping, access control, and version compatibility & Usability, onboarding, correct roles, documentation quality, and prevention of informal off-platform agreements \\
\bottomrule
\end{tabularx}
\end{table}

The dual-access model therefore supports both enterprise integration and inclusive participation. Large companies can connect BWLS to established HR and ERP workflows, while smaller organizations can participate through lightweight tools without losing the governance guarantees required for legally and ethically sensitive workforce leasing. This design choice is essential for positioning BWLS as a socio-technical platform rather than merely an enterprise plugin or an unregulated labor marketplace.

\newpage
\section{Governance-by-Design Architecture}

\subsection{Architectural Principle}

The central architectural principle is \emph{local control with minimal central coordination}. Each participating company runs a local connector or platform node, while personal profile data and detailed employee records remain under the control of the employer and the employee. This principle aligns enterprise-architecture concerns with privacy-by-design and platform-governance requirements \cite{OpenGroup2022ArchiMate,ISO42010,GDPR2016,EDPB2020DPbDD,Tiwana2010PlatformEvolution}.

This design reduces the risk of creating a centralized repository of sensitive employee information. It also aligns with data minimization and purpose limitation. However, decentralization does not remove the need for governance. Local nodes must still implement consistent logging, authentication, access control, encryption, and policy enforcement.

\begin{figure}[H]
\centering
\includegraphics[width=0.98\textwidth]{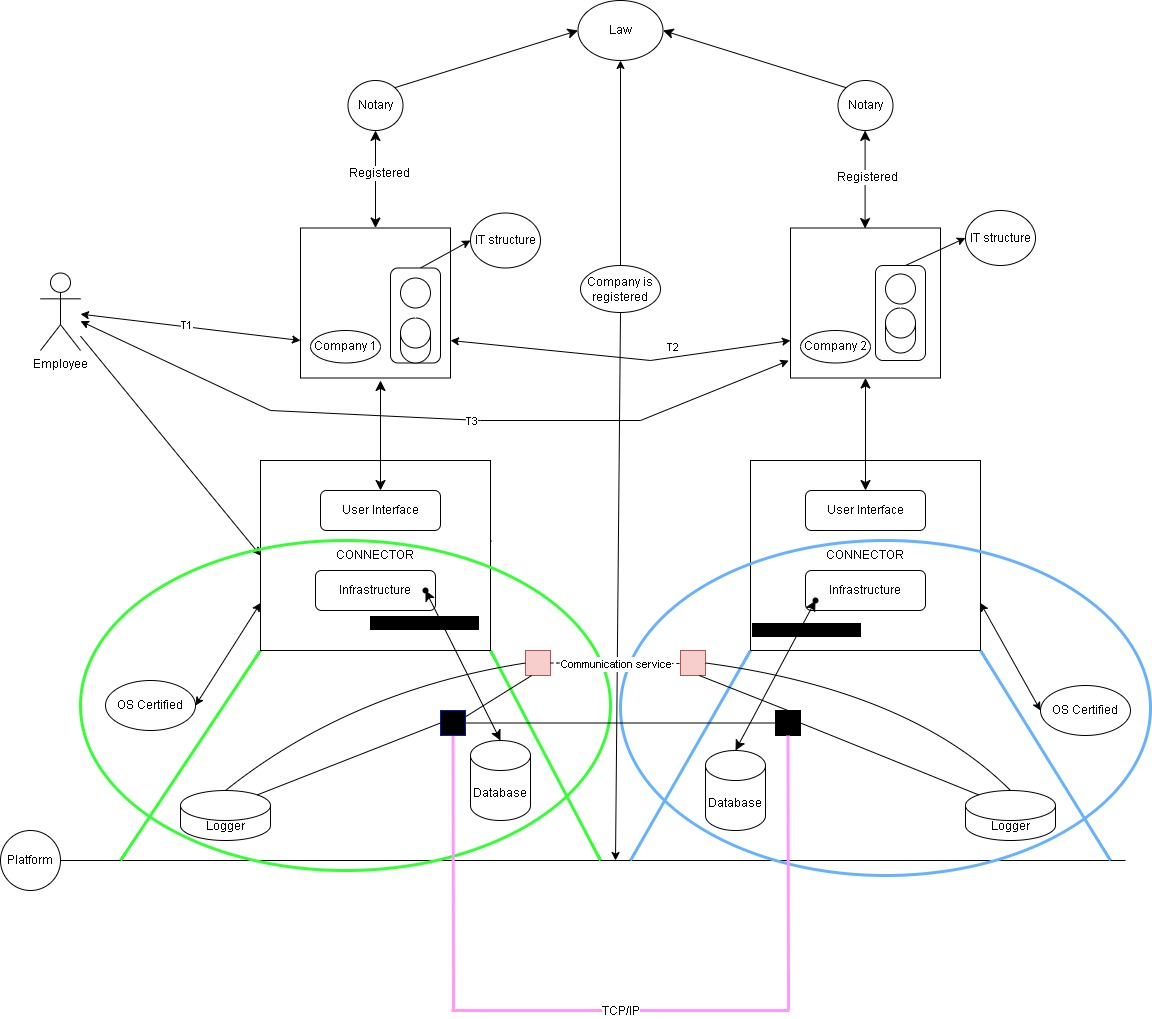}
\caption{Finalized platform sketch showing two company environments, local connectors, databases, loggers, a communication service, and TCP/IP-based coordination. The figure captures the design idea of distributed company nodes connected through a governed platform layer.}
\label{fig:loan_platform}
\end{figure}

\subsection{Core Components}

\begin{table}[H]
\centering
\caption{Governance responsibilities of core BWLS components}
\label{tab:components}
\small
\begin{tabularx}{\textwidth}{L{3.2cm}Y}
\toprule
\textbf{Component} & \textbf{Governance responsibility} \\
\midrule
Company connector & Provides secure access to the platform from the company's IT environment, integrates with ERP/HR systems, and enforces local policies. \\
\addlinespace
Local database & Stores detailed employee profiles, approvals, access logs, and company-specific records under local control. \\
\addlinespace
Central coordination layer & Stores minimal non-personal availability and transaction metadata needed for matching, request routing, licensing, and audit references. \\
\addlinespace
User interface & Enables HR, managers, and employees to manage profiles, requests, consent, communication, and history according to roles. \\
\addlinespace
Communication service & Supports secure negotiation between companies after mutual interest, with traceability and policy-aware logging. \\
\addlinespace
Contract management module & Generates, stores, signs, and archives leasing agreements, including terms, responsibilities, payment logic, and dispute procedures. \\
\addlinespace
Logger/audit service & Records material actions such as profile creation, consent, approval, interest expression, contract updates, and final agreement. \\
\bottomrule
\end{tabularx}
\end{table}

\subsection{Data Governance and Employee Rights}

Data governance in BWLS follows privacy-by-design and contextual-integrity principles: employee data should be minimized, purpose-bound, access-controlled, and disclosed only in appropriate transaction contexts \cite{GDPR2016,EDPB2020DPbDD,Nissenbaum2004Privacy,Spiekermann2009EngineeringPrivacy}. The employee profile should separate personal identifiers from skill, availability, and preference metadata. Before a profile becomes visible outside the original company, the employee and the employer should both approve the listing. The system should also provide a leasing history, feedback mechanisms, and the ability to inspect what information has been shared.

Key employee-facing rights in the framework include:

\begin{itemize}
\item informed consent before profile listing and before disclosure of identifiable information;
\item control over which qualifications, skills, preferences, and availability windows are visible;
\item access to a history of requests, assignments, feedback, and contracts;
\item ability to report issues with working conditions, payment, or treatment;
\item access to explanations for why a profile was selected, rejected, paused, or removed.
\end{itemize}

These rights are not merely policy statements. They must be implemented in the user interface, data model, and audit logs.

\subsection{Security Controls}

BWLS requires security controls at several layers. Transport security protects communication between nodes. Authentication and two-factor authentication restrict access to authorized users. Role-based access ensures that employees, HR managers, administrators, and external partners see only appropriate information. Audit trails support accountability, while retention limits prevent indefinite storage of sensitive records.

Docker-based deployment can support environment consistency and modular updates \cite{Merkel2014Docker}. However, containerization is not a complete security solution. The surrounding infrastructure must still handle patching, secrets management, network configuration, backup policy, vulnerability scanning, and incident response.

\section{Business Model and Value Network}

The business model analysis uses the Business Model Canvas and e3value perspective to connect platform sustainability with actor-level value exchanges and governance dependencies \cite{Osterwalder2010BusinessModel,Gordijn2001E3Value}.

\subsection{Business Model Canvas Summary}

BWLS can generate value for multiple actors. Companies with surplus talent monetize idle capacity and maintain employee engagement. Companies needing talent obtain temporary skills without permanent hiring. Employees gain project variety and professional development opportunities, provided that participation is voluntary and fair. ERP providers gain an additional service layer for their customers.

\begin{table}[H]
\centering
\caption{Condensed business model canvas for BWLS}
\label{tab:bmc}
\small
\begin{tabularx}{\textwidth}{L{3.2cm}Y}
\toprule
\textbf{Canvas element} & \textbf{BWLS interpretation} \\
\midrule
Key partners & ERP providers, HR consultancies, legal/compliance experts, technology partners, cybersecurity specialists. \\
\addlinespace
Key activities & Connector development, platform operation, customer support, legal template maintenance, compliance monitoring, ecosystem management. \\
\addlinespace
Key resources & Platform infrastructure, connectors, documentation, legal templates, support team, developer ecosystem, audit infrastructure. \\
\addlinespace
Value propositions & Idle workforce monetization, temporary access to skills, employee development opportunities, ERP ecosystem extension. \\
\addlinespace
Customer relationships & Self-service portal, dedicated support, onboarding assistance, dispute resolution, developer community. \\
\addlinespace
Channels & ERP marketplaces, direct enterprise sales, HR networks, industry events, online marketing, developer portal. \\
\addlinespace
Customer segments & ERP-using companies, non-ERP small and medium enterprises, employees, ERP providers, HR service partners. \\
\addlinespace
Cost structure & R\&D, infrastructure, customer support, cybersecurity, legal compliance, sales, documentation, integration maintenance. \\
\addlinespace
Revenue streams & License fees, subscription fees, transaction fees, installation services, training, verification services, premium analytics. \\
\bottomrule
\end{tabularx}
\end{table}

\subsection{Governance Implications of the Business Model}

The BWLS business model has direct governance implications. License and subscription fees support long-term maintenance, while transaction fees create incentives around successful workforce-leasing agreements. However, the platform should not reward transaction volume alone. A match should only count as successful when employee consent, contractual validity, data minimization, approval traceability, and agreement documentation are satisfied.

This means that governance must be embedded into both the architecture and the revenue logic. BWLS should operate as a regulated socio-technical coordination infrastructure rather than as an unstructured labor marketplace. ERP providers, HR consultants, legal experts, cybersecurity specialists, companies, and employees all contribute to this ecosystem, but the platform must coordinate them through auditable workflows, certified connectors, clear permissions, and enforceable process rules.

A sustainable BWLS model therefore needs to balance economic value with employee autonomy, legal compliance, and organizational accountability. This connects the Business Model Canvas to the e3value-inspired analysis, where the focus shifts from business components to concrete value exchanges between actors \cite{Osterwalder2010BusinessModel,Gordijn2001E3Value,Tiwana2010PlatformEvolution,deReuver2018DigitalPlatforms}.

\subsection{Adoption and Sustainability Considerations}

The adoption of BWLS depends on whether organizations perceive the platform as trustworthy, useful, and manageable within their existing work practices. Large companies may value ERP integration, compliance alignment, and auditability, while smaller organizations may prioritize low setup effort, clear documentation, and affordable access. Therefore, the platform must support different adoption paths without fragmenting the underlying governance model.

Sustainability also depends on continuous platform maintenance. Connectors must remain compatible with changing ERP systems, legal templates must be updated when regulations evolve, and security mechanisms must be monitored over time. If these elements are neglected, the platform could lose trust even if the original architecture is technically sound. Long-term operation therefore requires not only software development, but also support processes, certification routines, and governance updates.

For BWLS, adoption and sustainability are closely connected. A platform that is too complex may fail to attract users, while a platform that is too lightweight may fail to protect employees, companies, and sensitive data. The business model must therefore fund the infrastructure, support, and compliance work required to keep the system reliable, inclusive, and accountable over time \cite{Tiwana2010PlatformEvolution,deReuver2018DigitalPlatforms,GDPR2016,EDPB2020DPbDD}.

\subsection{e3value-Inspired Value Exchanges}

The e3value perspective helps identify whether the business model distributes value coherently. BWLS includes five main actors: companies with surplus talent, companies needing talent, employees, ERP providers, and the BWLS operator. Value objects include platform access, leased employee time and skills, subscription fees, transaction fees, integration services, and contractual assurance.

\begin{table}[H]
\centering
\caption{Illustrative value exchanges in the BWLS ecosystem}
\label{tab:e3value}
\small
\begin{tabularx}{\textwidth}{L{3.4cm}L{3.4cm}Y}
\toprule
\textbf{Source actor} & \textbf{Target actor} & \textbf{Value exchange} \\
\midrule
Lessor company & BWLS & Subscription, transaction fee, and employee availability information. \\
\addlinespace
BWLS & Lessor company & Matching infrastructure, auditability, contract workflow, and reduced idle workforce cost. \\
\addlinespace
Lessee company & BWLS & Search, request, subscription, or transaction payments. \\
\addlinespace
BWLS & Lessee company & Access to temporary qualified talent, negotiation tools, documentation, and traceable agreements. \\
\addlinespace
Employee & Lessor / lessee & Time, skills, consent, and assignment participation. \\
\addlinespace
Lessor / lessee & Employee & Salary continuity, professional development, feedback, and assignment conditions. \\
\addlinespace
ERP provider & BWLS & Integration partnership, distribution channel, or marketplace relationship. \\
\addlinespace
BWLS & ERP provider & Connector functionality and added value for ERP customers. \\
\bottomrule
\end{tabularx}
\end{table}

From a design perspective, the e3value analysis clarifies that BWLS is only viable if value exchanges remain reciprocal and visible to all actors. Lessor companies should not only reduce idle workforce costs, but also preserve employee development and retention. Lessee companies should gain temporary skills without bypassing legal responsibility. Employees should not be treated as transferable resources alone, but as active participants whose consent, conditions, and feedback shape the transaction. ERP providers and the BWLS operator gain ecosystem value only if the platform maintains trust, interoperability, and compliance. The value network therefore reinforces the central argument of this paper: technical architecture, business incentives, and governance rules must be designed together rather than treated as separate layers \cite{Gordijn2001E3Value,Osterwalder2010BusinessModel,Tiwana2010PlatformEvolution}.

The value model also reveals governance dependencies. If revenue depends strongly on transaction volume, the platform may be incentivized to maximize matches even when employee consent or quality control requires caution. Governance mechanisms must therefore counterbalance commercial incentives.

\newpage
\section{Legal, Ethical, and Operational Risk Analysis}

The Balanced Workforce concept introduces legal, ethical, and operational risks. A responsible framework should make these risks explicit rather than presenting the platform as an automatically beneficial innovation. The risk analysis combines labor-law, privacy, platform-governance, and socio-technical perspectives because workforce leasing affects both organizational efficiency and employee rights \cite{EU2008TemporaryAgencyWork,DeStefano2016JustInTime,GDPR2016,Kellogg2020AlgorithmsAtWork,Wood2019GoodGigBadGig}.

\begin{table}[H]
\centering
\caption{Risk matrix and governance controls}
\label{tab:risk_matrix}
\small
\begin{tabularx}{\textwidth}{L{3.0cm}Y Y}
\toprule
\textbf{Risk area} & \textbf{Risk} & \textbf{Governance control} \\
\midrule
Labor law & Assignments may violate local labor rules, working-time limits, equal-treatment requirements, or social-insurance obligations. & Jurisdiction-specific legal templates, expert review, mandatory contract fields, working-condition checks, and approval gates. \\
\addlinespace
Data protection & Employee profiles, availability, skills, and feedback may expose personal information or enable re-identification. & Data minimization, anonymized listings, local profile storage, purpose limitation, retention limits, and data protection impact assessment. \\
\addlinespace
Consent and autonomy & Employees may feel pressured to participate or may not understand the consequences of being listed. & Explicit opt-in, clear consent screens, withdrawal rules, employee-facing explanations, and independent complaint channels. \\
\addlinespace
Fairness and bias & Profile ranking, selection, and feedback may produce discrimination or unequal access to opportunities. & Transparent criteria, audit logs, bias monitoring, appeal procedures, and limits on sensitive attributes. \\
\addlinespace
Conflict of interest & Leasing to competitors may expose confidential information, trade secrets, or strategic knowledge. & Conflict checks, confidentiality clauses, restricted assignment categories, and employer approval workflows. \\
\addlinespace
Intellectual property & Work created during an assignment may create disputes over ownership. & Contract clauses defining IP ownership, licensing, deliverables, and responsibility boundaries. \\
\addlinespace
Quality control & Poor skill matching may harm projects and employee reputation. & Verification of skills, structured feedback, project-fit review, trial periods, and dispute-handling workflow. \\
\addlinespace
Security & Connectors and central services may become attack surfaces. & Secure development, encryption, 2FA, role-based access, vulnerability scanning, incident response, and logging. \\
\addlinespace
Scalability and reliability & Growth may overload matching, logging, communication, or contract workflows. & Modular architecture, monitoring, queues, load testing, backup strategy, and service-level targets. \\
\bottomrule
\end{tabularx}
\end{table}

\subsection{Contractual Governance}

The contractual layer is central. A leasing agreement should specify assignment duration, supervision, payment responsibilities, salary continuity, insurance, working conditions, confidentiality, intellectual property, liability, termination, feedback, data processing, and dispute resolution. The platform should not allow a transaction to reach T3 unless mandatory fields are completed and approved by authorized parties.

\subsection{Ethical Governance}

The main ethical risk is commodification: employees may be treated as movable resources rather than as people with autonomy, preferences, and professional goals. BWLS must therefore provide employee-facing controls and not only employer-facing efficiency tools. Ethical governance also requires transparency about selection, feedback, reputation, and assignment consequences. Ratings should be handled carefully because they can become long-term reputational infrastructure.

\subsection{Operational Governance}

Operationally, BWLS depends on integration reliability. ERP connectors, standalone applications, secure communication, and contract workflows must work across heterogeneous company environments. The architecture should therefore prioritize modularity and clear failure states. If a transaction cannot be logged, if a consent record is missing, or if an identity check fails, the platform should pause the transaction rather than allowing untraceable off-platform completion.

\section{Proposed Evaluation Agenda}

Because the present paper is a framework contribution, evaluation should follow staged design-science validation rather than immediate real-world deployment \cite{Hevner2004DesignScience,Gregor2006DesignScience}.

\begin{table}[H]
\centering
\caption{Proposed evaluation plan for future BWLS research}
\label{tab:evaluation}
\small
\begin{tabularx}{\textwidth}{L{3.2cm}Y Y}
\toprule
\textbf{Evaluation stage} & \textbf{Purpose} & \textbf{Possible evidence} \\
\midrule
Legal expert review & Assess whether contract templates, role definitions, and data flows are legally plausible. & Legal checklist, compliance report, jurisdiction-specific risk notes. \\
\addlinespace
Stakeholder workshop & Test whether employees, HR, managers, and compliance actors understand and accept the process. & Interview notes, workshop artifacts, perceived risks, design changes. \\
\addlinespace
Prototype usability test & Evaluate whether users can complete T0--T3 tasks and understand consent and disclosure states. & Task completion, errors, perceived ease of use, comprehension questions. \\
\addlinespace
Security and privacy review & Identify vulnerabilities and privacy risks in connectors, APIs, storage, and logging. & Threat model, DPIA, penetration-test notes, data-flow diagrams. \\
\addlinespace
Simulation study & Test matching, transaction load, logging completeness, and failure recovery without real employees. & Synthetic transaction logs, latency, error handling, scalability measures. \\
\addlinespace
Pilot deployment & Carefully evaluate a limited real-world trial with volunteer companies and employees. & Assignment outcomes, disputes, satisfaction, compliance events, audit completeness. \\
\bottomrule
\end{tabularx}
\end{table}

The evaluation should not only measure platform efficiency. It should also measure worker understanding, consent quality, perceived fairness, privacy comfort, trust in audit mechanisms, and ease of dispute reporting.

\section{Discussion}

BWLS sits between enterprise architecture, labor governance, and platform design, making it a socio-technical system whose risks cannot be evaluated through technical matching performance alone \cite{Baxter2011SocioTechnical,Mumford2006Story,Tiwana2010PlatformEvolution,deReuver2018DigitalPlatforms}.

The most important design lesson is that matching is not the core difficulty. Matching can be implemented through relatively standard search, filtering, and profile mechanisms. The harder challenge is governing the conditions under which matching is allowed to proceed. Who can list an employee? What data is visible at each stage? How is consent recorded? When does anonymity end? Who owns work products? What happens if the assignment fails? How are disputes escalated? These questions define the socio-technical nature of the platform.

A second lesson is that decentralization and local connectors can reduce some privacy risks but also create coordination challenges. If every company hosts its own connector, the platform must ensure interoperability, consistent logging, and consistent policy enforcement. If the central layer stores too little information, dispute resolution may become difficult. If it stores too much, privacy risk increases. The design must therefore balance minimality and accountability.

A third lesson concerns business incentives. Subscription, licensing, transaction fees, and premium services can sustain the platform, but revenue logic should not override worker protection. The framework should therefore include governance metrics alongside business metrics. Success should not be measured only by the number of assignments, but also by consent quality, dispute resolution speed, employee satisfaction, privacy incidents, and fair access.

\subsection{Current Regulatory and Algorithmic-Management Context}

Recent regulatory and policy developments strengthen the relevance of BWLS as a governance-by-design problem. The EU Platform Work Directive shows that platform-mediated labor is increasingly treated as a domain where working conditions, transparency, automated decision-making, and human oversight require explicit regulation \cite{EU2024PlatformWorkDirective}. Although BWLS is not a gig-work platform in the narrow sense, it shares several governance concerns with platform work: profile visibility, matching logic, access to opportunities, monitoring, feedback, dispute handling, and the risk that software-mediated coordination may reshape employment relationships.

The relevance of these concerns is reinforced by recent work on algorithmic management. The OECD defines algorithmic management as the use of software, including AI-enabled tools, to fully or partially automate managerial tasks, and reports new evidence on how such systems are used in workplaces \cite{Milanez2025AlgorithmicManagement}. For BWLS, this matters because even a seemingly neutral matching platform can become managerial infrastructure once it influences which employees are visible, which profiles are prioritized, which assignments are recommended, and how feedback affects future opportunities.

A recent multi-stakeholder study on algorithmic management and workplace scheduling further shows that legal compliance cannot be reduced to static software rules alone. Lynn et al. argue that regulating algorithmic management involves rule operationalization, software use, and enforcement, and that stakeholders such as regulators, workers, advocates, attorneys, and managers may experience the same software system differently \cite{Lynn2025RegulatingAM}. This insight is directly relevant for BWLS because the platform would not merely store data or route requests; it would mediate legally and socially meaningful decisions across organizations.

Therefore, future BWLS implementations should avoid opaque ranking or automated assignment decisions. If matching, recommendation, prioritization, analytics, or reputation features are introduced, they should remain explainable, contestable, auditable, and subordinate to employee consent and contractual review. The platform should also separate decision support from decision authority: software may help identify potential matches, but employees, employers, and legally responsible actors must remain able to review, reject, appeal, or pause transactions before an assignment becomes binding.

This regulatory context also strengthens the paper's central architectural claim. Governance cannot be added only through policy documents after the platform is deployed. It must be represented in system states, permissions, logs, user interfaces, contract workflows, data-retention rules, and escalation mechanisms. BWLS should therefore be evaluated not only as a workforce-matching system, but as a socio-technical governance infrastructure whose design choices affect autonomy, transparency, accountability, and fair access to work opportunities.

\section{Limitations}

This paper has several limitations. First, it is based on a conceptual design and seminar project rather than a deployed platform. Therefore, it cannot claim real-world adoption, legal validity in a specific jurisdiction, economic profitability, or measured labor-market impact.

Second, the architecture remains high-level. Implementation details such as API schemas, authentication protocols, identity federation, encryption design, log retention, container orchestration, and integration adapters require further specification.

Third, the legal analysis is indicative rather than legal advice. Workforce leasing is highly jurisdiction-dependent, and any deployment would require professional legal review.

Fourth, the paper does not include interviews with employees, HR managers, legal experts, or ERP providers. Such stakeholder data is necessary before deployment.

Finally, the platform concept may be appropriate only for specific industries, employment relationships, and jurisdictions. It should not be generalized to all forms of labor intermediation.

\section{Future Work}

Future work should proceed in five directions. First, the governance model should be refined through interviews and workshops with employees, HR managers, legal experts, enterprise architects, and data-protection officers. Second, a clickable prototype should test consent flows, profile visibility, request handling, negotiation, and contract approval. Third, a data protection impact assessment should map personal data, risks, access rights, and controller/processor roles.

Fourth, the enterprise architecture should be specified more formally using ArchiMate viewpoints to clarify business roles, application services, data objects, technology nodes, and cross-company interfaces. Fifth, the business model should be evaluated with realistic cost assumptions, including legal support, cybersecurity, maintenance, integration support, and customer acquisition.

The original project also identified blockchain-based contract signing, analytics dashboards, and global expansion as future directions. These should be treated carefully. Blockchain can support traceability, but may conflict with privacy and deletion requirements. Analytics dashboards may support planning, but risk surveillance and biased evaluation.

A further direction is to evaluate BWLS against emerging rules and expectations for algorithmic management. If future versions include matching scores, recommendation systems, prioritization logic, analytics dashboards, or reputation mechanisms, these features should be reviewed for explainability, contestability, bias, and worker impact. This requires more than technical testing: future work should define governance metrics for consent quality, transparency, appealability, dispute resolution, and fairness of opportunity. Such metrics would help ensure that BWLS remains a decision-support and coordination infrastructure rather than becoming an opaque system for automated workforce allocation \cite{EU2024PlatformWorkDirective,Milanez2025AlgorithmicManagement,Lynn2025RegulatingAM}.

\section{Conclusion}

This paper presented Balanced Workforce as a governance-by-design framework for privacy-preserving inter-firm workforce leasing. The framework responds to a practical organizational problem: companies may simultaneously experience temporary skill shortages and underutilized workforce capacity. A digital platform can connect these needs only if governance is embedded into the architecture.

The proposed BWLS framework uses local company connectors, minimal central coordination, anonymized skill listings, secure communication, contract management, audit trails, role-based access, and employee consent mechanisms. It also connects the technical architecture to a business model canvas, e3value-inspired value exchanges, and a risk matrix covering legal, ethical, and operational challenges.

The paper does not claim that BWLS is deployment-ready. Instead, it provides a structured design artifact and research agenda. Future work should evaluate the framework through legal review, stakeholder workshops, prototype testing, security analysis, and pilots. The central argument is that workforce sharing platforms should be designed not only for efficiency, but also for accountable, privacy-preserving, employee-aware governance.

\section{Project Repository and Materials Availability}

The project repository associated with this work is publicly available on GitHub:

\begin{center}
\small
\href{https://github.com/Ariton123/engineering-lab/tree/main/coursework/master/design-of-governance-in-socio-technical-information-systems}
{\texttt{Balanced Workforce project repository}}
\end{center}

The repository contains course project materials related to the Balanced Workforce Leasing Service, including design documentation, architectural diagrams, and supporting artifacts developed during the Design of Governance in Socio-Technical Information Systems course.

The repository should be interpreted as a project artifact and design-materials repository rather than as a production-ready software release. The paper presents a conceptual and architectural framework, while future work may extend the repository with executable prototypes, deployment scripts, evaluation material, and implementation details.


\end{document}